\documentclass[a4paper, 12pt]{article}

\usepackage{dsfont}
\usepackage{setspace}
\usepackage{cancel}
\usepackage{url}
\usepackage{xcolor}
\usepackage{bm,bbm}
\usepackage{fancyhdr}
\usepackage{empheq}
\usepackage{float}

\usepackage[colorlinks=true]{hyperref}
\hypersetup{
linkcolor=blue,
citecolor=blue,
urlcolor=blue
}

\usepackage{epsfig}
\usepackage{eucal}
\usepackage{verbatim}
\usepackage{graphicx}
\usepackage{amsmath}
\usepackage{amsfonts}
\usepackage{amssymb}
\usepackage{numprint}
\usepackage{soul}
\usepackage{cite}

\advance\oddsidemargin-0.65in
\definecolor{gray}{rgb}{0.4,0.4,0.4}

\begin{document}

\thispagestyle{fancy}
\lhead{}
\rhead{}
\renewcommand{\headrulewidth}{0pt}
\renewcommand{\footrulewidth}{0pt}
\fancyfoot[C]{\footnotesize \textcolor{gray}{}}

\title{Upstreamness and downstreamness in input-output analysis from local and aggregate information}

\author{Silvia Bartolucci\footnote{Dept. of Computer Science, University College London, 66-72 Gower Street WC1E 6EA London (UK). Centre for Financial Technology, Imperial College Business School, South Kensington SW7 2AZ London (UK).}, Fabio Caccioli \footnote{Dept. of Computer Science, University College London, 66-72 Gower Street WC1E 6EA London (UK). Systemic Risk Centre, London School of Economics and Political Sciences, WC2A 2AE, London (UK). London Mathematical Laboratory, 8 Margravine Gardens, London WC 8RH (UK).}, Francesco Caravelli\footnote{T-Division (Center for Nonlinear Studies and T4), Los Alamos National Laboratory, Los Alamos NM 87545 (USA).}, Pierpaolo Vivo\footnote{Dept. of Mathematics, King's College London, Strand WC2R 2LS London (UK).}}

\date{}
\maketitle
\begin{abstract}
Ranking sectors and countries within global value chains is of paramount importance to estimate risks and forecast growth in large economies. However, this task is often non-trivial due to the lack of complete and accurate information on the flows of money and goods between sectors and countries, which are encoded in Input-Output (I-O) tables. In this work, we show that an accurate estimation of the role played by sectors and countries in supply chain networks can be achieved without full knowledge of the I-O tables, but only relying on local and aggregate information, e.g., the total intermediate demand per sector. 
Our method, based on a rank-$1$ approximation to the I-O table, shows consistently good performance in reconstructing rankings (i.e., upstreamness and downstreamness measures for countries and sectors) when tested on empirical data from the World Input-Output Database. Moreover, we connect the accuracy of our
approximate framework with the spectral properties of the I-O tables, which ordinarily exhibit relatively large spectral gaps.
Our approach provides a fast and analytical tractable framework to rank constituents of a complex economy without the need of matrix inversions and the knowledge of finer intersectorial details.
\end{abstract}
\vspace{1.3cm}
    
\section{Introduction}
The introduction of Input-Output (I-O) analysis as a fundamental tool to analyze the inter-relationship between economic sectors of a country was pioneered by W. Leontief, who proposed the construction of the first I-O tables for the United States for the years 1919 and 1929 \cite{Leontief1936,Leontiefbook}. An I-O table summarizes how the products (outputs) of a given industry or economic sector are used as input to other industries or sectors within the same, or different, economies (for instance, in the case of Import/Export exchanges with other countries) \cite{handbook}.
Understanding the structure and relevance of industrial sectors and countries within the so-called \textit{global value chains }(GVCs), encompassing the different stages of the production process across different countries, is of central importance \cite{antras2012econometrica}. To achieve this, a number of indicators and measures have been devised that characterize the relative positioning of industries and economic sectors in the economy. These rely on the calculation of the following technical object,
\begin{equation}
    G(A)=(\mathds{1}_N-A)^{-1}\ ,\label{eq:resolvent}
\end{equation}
the so-called {\em Leontief inverse} (or \emph{resolvent}) matrix. Here, $\mathds{1}_N$ is the $N\times N$ identity matrix (where $N$ is the number of industrial sectors) and $A$ is a sub-stochastic matrix with positive entries, which is related in a simple fashion to the original I-O table.
Notably, the upstreamness and downstreamness metrics proposed by Antr\'as, Chor and collaborators (see Sec. \ref{sec:defUD} for mathematical definitions) have become widely used and mainstream in recent years \cite{AntrasFally2012,Fally2012,Miller}. One of the main practical challenges of the Input-Output analysis lies in the accurate and reliable compilation of inter-sectorial I-O tables from which the matrix $A$ in formula \eqref{eq:resolvent} is derived. This issue is particularly felt at firm-level, where often only aggregate information is available \cite{bacilieri}.

The main contribution of our paper is to show that up-/downstreamness measures and similar resolvent-like metrics can be approximated with high accuracy even when possessing only {\em aggregate} and {\em local} information about the inter-sectorial dependencies encoded within the I-O table. In this case, the required information only amounts the row (or column) sums of the matrix $A$, representing the total intermediate
demand per industry (or the total value of all inputs required by each industry).

More specifically, we propose an approach rooted in complexity science that reconstructs the most likely matrix $A$ derived from I-O tables on the basis of limited/aggregated information, and uses this surrogate information to compute the Leontief inverse and related indicators (e.g., upstreamness and downstreamness). These indicators can be derived from the aggregate information available in a fast -- as this procedure does not require to perform a full matrix inversion-- and accurate way. Moreover, in this work we connect the accuracy of our approximate framework with the spectral properties of the I-O tables.

\subsection{Related literature}
There is a vast literature concerning input-output models and how inaccuracies and noise in I-O tables may affect the determination of the relative ranking of industrial sectors and countries within the economy.
One strand focuses on the accuracy of the \emph{empirical} input-output matrix denoted by $A_{emp}$ with respect to the \emph{true} matrix $A_{true}$. The main question is about how errors  occurring in the compilation of the Input/Output tables propagate and affect measurements and predictions based on nonlinear functions of $A_{emp}=A_{true}+H$ (for instance, the Leontief Inverse $(\mathds{1}_N-A_{emp})^{-1}$), where $H$ encodes the stochastic sources of error. Compiling the entries of the matrix $A_{emp}$ is subject to many issues, for instance the difficulty in sampling and surveying firms and flows of goods with great accuracy \cite{KopJansen1994,KopJansen1990}. This has provided the motivation to study stochastic models for the input-output analysis.

Evans \cite{Evans1954} and Quandt \cite{Quandt1958} are among the first to look at this problem by constructing random models. Evans \cite{Evans1954} assumed that the error matrix $H$ had only one non-zero row and that the errors could be propagated on a row-by-row basis. Quandt \cite{Quandt1958} assumed that the errors $H_{ij}$ on the matrix elements are independent and normally distributed with mean zero, solved the error propagation problem for a small-size system (e.g. $2\times 2$), and determined the confidence intervals on the expected Leontief Inverse. Later, Simonovits \cite{Simonovits1975} deduced the fundamental inequality $\langle (\mathds{1}_N-A_{emp})^{-1}\rangle_H\geq (\mathds{1}_N-\langle A_{emp}\rangle_H)^{-1}$, where the average is taken with respect to independent matrix elements of $H$. This inequality circumvents the problem of inverting the matrix $\mathds{1}_N-A_{emp}$, which has so far been one of the major and long-standing theoretical challenges. 

One of the first comprehensive theoretical studies of stochastic input-output models is due to West \cite{West1986}. His starting point is a random matrix $H$, of which the expected value and the standard error of all the elements are known, with the aim to provide approximating formulas for the expected value and the standard errors of the Leontief Inverse in terms of these known quantities. Some of the assumptions (for instance, that the errors $H_{ij}$ be independent and normally distributed) are however not realistic or plainly incompatible with the sub-stochasticity constraint, and only lead to a closed-form solution for the mean and variances of the deviations from the ``true" matrix under very restrictive choices for the variances of the errors in $H$.

More recently, this approach has been re-evaluated by Kogelschatz \cite{Kogelschatz2007} - who assumed that the $a_{ij}$ are Beta-distributed and derived estimates for the elements of the Leontief Inverse  -
and Kozicka \cite{Kozicka2019} - who postulated more realistic distribution for the matrix entries, but provided explicit formulae only for small-size systems.  

Within the empirical literature, a number of studies have been also undertaken to characterize the regional inter-sectorial dependence of industries and to discuss the challenges of reconstructing regional data from national accounts and surveys \cite{reviewregional}.

Given the practical difficulties associated with compiling input-output tables, especially at the regional level, earlier scholars devised ``shortcut" methods to estimate the Leontief inverse from incomplete or unreliable information, or even foregoing I-O tables altogether. Katz and Burford \cite{Katz, burford} derived a formula under the assumption that the matrix $A$ is uniformly drawn from the set of sub-stochastic matrices, and under the rather questionable technical condition that the covariance between the entries of the matrix and the output multipliers be null. Their work hinges on an earlier formula empirically derived by Drake \cite{drake}. The general approach based on finding ``shortcuts" and foregoing a painstaking compilation of I-O tables was criticized on both technical and conceptual grounds \cite{Phibbs,alive, resurrection, requiem} before this line of investigation was dropped and even ignored altogether in the subsequent related literature.

The Leontief inverse and the associated indicators have also been looked at through the prism of complexity and network science. Cerina et al. \cite{Cerina2015} analyzed the properties of the (global and regional) network of industries in different economies reconstructing the monetary goods flows (edges) using the input-output matrix. McNerney et al. \cite{McNerney2018} used average national output multipliers to predict future economic growth and price changes. In \cite{moran2019}, a model for the propagation and amplification of idiosyncratic shocks along the input-output network is provided. In \cite{jensen2017}, a network analysis of the World Input-Output Data set is undertaken to analyze the temporal interdependence between countries and industrial sectors.

In recent years the interest in input-output models has grown steadily \cite{Carvalho2012}, also in view of a rather compelling connection to models of complexity and networks \cite{acemoglu,jensen2017}. Moreover, many of these ideas can in principle be extended to more general sector-product spaces, which saw many uses for the study of the connection between complexity measures, productivity and economic growth \cite{hidalgo1,hidalgo2,pietronero1,pietronero2}. 

In the following section, we will focus on the works by Antr\`{a}s and Chor \cite{antras2012econometrica}, Fally et al. \cite{Fally2012} and Miller et al. \cite{Miller}, where different incarnations of the so-called upstreamness and downstreamness measures have been first proposed.
An early example of  a direct application of those measures for the analysis of empirical data on global value chains can be found in \cite{Antras2018}, now used in multiple contexts \cite{carbon,shock}.

\section{Definition of Upstreamness and Downstreamness}\label{sec:defUD}
\begin{figure}[htb!]
    \centering
    \includegraphics[width=0.7\textwidth]{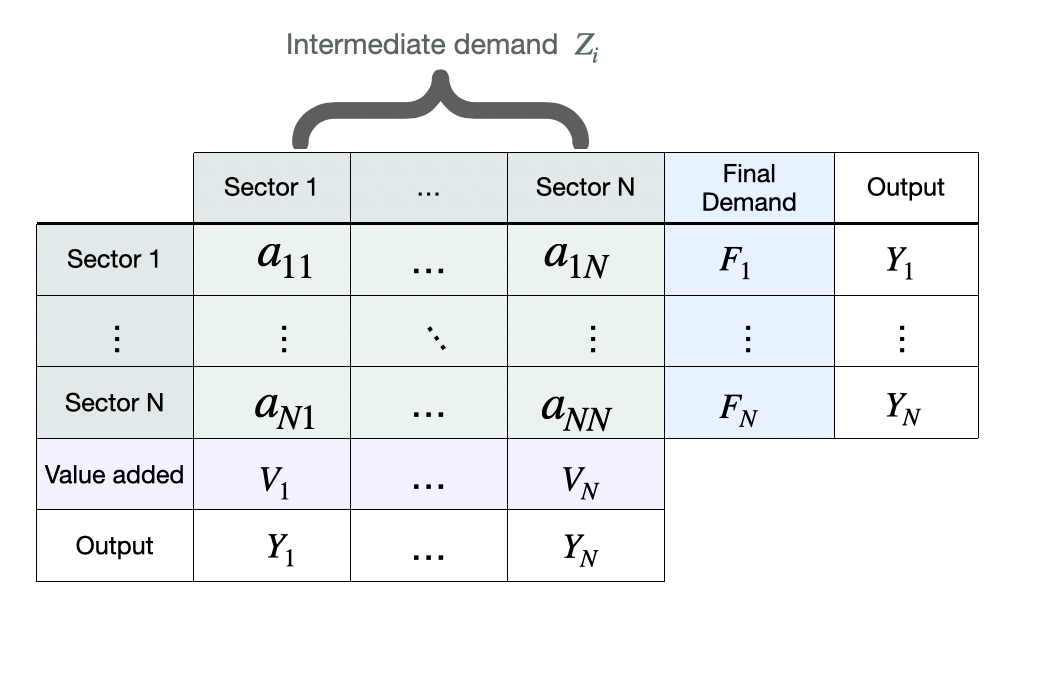}
    \caption{Scheme of the structure of a single-country Input-Output table \cite{wiotdataset,handbook, suganuma}.}
    \label{fig:ioscheme2}
\end{figure}

Antr\`{a}s et al. \cite{antras2012econometrica} considered a closed economy of $N$ industries. For each industrial sector $i= 1, \dots, N$ we indicate the value of gross output with $Y_i$ and the total intermediate demand (i.e., the use of the output of an industry as a final good) with $F_i$. Then the following equality holds in Input-Output tables:
\begin{eqnarray}
Y_i&= F_i +Z_i = F_i+\sum_{j=1}^N a_{ij}=\label{eq:iteration1} \\
&=F_i +\sum_{j=1}^N d_{ij}Y_j \label{eq:iteration}\ ,
\end{eqnarray}
with $Z_i = \sum_{j=1}^N d_{ij}Y_j$ corresponding to the output of industry $i$ used as intermediate input to other industries ({\em intermediate demand}) as shown in the scheme in Fig. \ref{fig:ioscheme2}. In \cite{antras2012econometrica}, $\{d_{ij}\}$ corresponds to the dollar amount of sector $i$'s output used to produce one dollar worth of sector $j$'s output and it is related to the matrix $A$ via the relationship $d_{ij}Y_j = a_{ij}$. The final demand, as detailed in Sec. \ref{Dataset}, comprises  contributions from different factors including, among others, the final consumption expenditure by households and government, and exports.

Iterating the identity Eq. \eqref{eq:iteration1} within Eq. \eqref{eq:iteration}, one obtains an infinite sequence of contributions, each representing the use of sector $i$'s output at different levels within the value chain \cite{handbook}
\begin{equation}
  Y_i = F_i + \sum_{j=1}^N d_{ij}F_j + \sum_{j=1}^N \sum_{k=1}^N d_{ik}d_{kj}F_j +\ldots\ . \label{eq:recursion}
\end{equation}
We can finally rewrite Eq. \eqref{eq:recursion} as follows
\begin{equation}
    {\bm Y} = [\mathds{1}_N-D]^{-1}{\bm F}
\end{equation}
using $\sum_{k\geq0} D^k=[\mathds{1}_N-D]^{-1}$. In this case, $\mathds{1}_N$ is the $N\times N$ identity matrix, $D=(d_{ij})$ contains each sector's output in dollar values and $\bm F$ is the vector of final demands.
 Antr\`{a}s et al. \cite{antras2012econometrica} hence proposed the following measure of upstreamness of the $i$-th industrial sector
\begin{equation}
U_{1i}= 1 \cdot \frac{F_i}{Y_i} + 2 \cdot\frac{\sum_{j=1}^N d_{ij}F_j}{Y_i} + 3 \cdot \frac{\sum_{j,k=1}^N d_{ik}d_{kj}F_j}{Y_i} + \ldots = \frac{([\mathds{1}_N - D]^{-2}{\bm F})_i}{Y_i} \ ,
\label{eq:up0}\end{equation}
where each term contributing to Eq. \eqref{eq:recursion} is weighted by their distance from final use and divided by the output of the sector $Y_i$.  The notation $(\cdot)_i$ is used to indicated the $i$-th component of the vector.
By construction, the terms of the sum that are further upstream in the value chain carry larger weight in the calculation of the upstreamness.
Inserting Eq. \eqref{eq:recursion} in Eq. \eqref{eq:up0}, we can rewrite the upstreamness as
\begin{equation}
  {\bm U_1} =  [\mathds{1}_N-A_U]^{-1}{\bm 1}_N \ , \label{eq:up1}
\end{equation}
where
\begin{equation}
A_U= Y^{-1}A = \begin{pmatrix} \frac{a_{11}}{Y_1} & \cdots & \frac{a_{1N}}{Y_1} \\
\vdots& \ddots & \vdots  \\
\frac{a_{N1}}{Y_N} & \cdots & \frac{a_{NN}}{Y_N}  \end{pmatrix}\  \label{eq:AU}
\end{equation}
and $Y ={\rm diag}(Y_1,\dots,Y_N)$. The vector ${\bm 1}_N$ is a column vector of $N$ ones. The matrix $A_U$ has non-negative elements, and in this convention it is row-substochastic, i.e., $\sum_{j}(A_U)_{ij}\leq 1 \ \forall j$.
 By construction $U_{1i}\geq 1$ and it is precisely equal to $1$ if no output of industry $i$ is used as input to other industries but it is only used to satisfy the final demand.

Later, Antr\`{a}s et al. \cite{AntrasFally2012} also established an equivalence between their upstreamness measure and a measure -- defined in a recursive fashion -- of the ``distance" of an industry from the final demand proposed independently by Fally et al. \cite{Fally2012}. Fally's upstreamness $U_2$ is defined as follows:
\begin{equation}
    U_{2i} = 1 + \sum_{j=1}^N\frac{d_{ij}Y_j}{Y_i}U_{2j} \ .
    \label{eq:fally1}
\end{equation}
The idea is that $\bm U_2$ aggregates information on the extent to which a sector in a given country produces goods that are sold directly to final consumers or that are sold to other sectors that themselves mainly sell  to final consumers.  Sectors selling a large share of their output to relatively upstream industries should be therefore considered to be more upstream themselves.
Using the fact that $d_{ij}Y_j = a_{ij}$ we obtain 
\begin{equation}
  {\bm U_2}= [\mathds{1}_N-A_U]^{-1}{\bm 1}_N\ ,
        \label{eq:fally2}
\end{equation}
where $A_U$ is defined in Eq. \eqref{eq:AU} as presented in \cite{AntrasFally2012}. 

On the input-side, there exists an analogous accounting identity stating that sector $i$’s total input $Y_i$ is equal to the value of its primary inputs (the so-called value added) $V_i$ plus its intermediate input purchased from all other sectors, namely
\begin{equation}
   Y_i= V_i +Z_i = V_i +\sum_{j=1}^N a_{ji}= V_i +\sum_{j=1}^N d_{ji}Y_j \ ,
\end{equation}
and 
\begin{equation}
  {\bm Y}= [\mathds{1}_N-D^T]^{-1}{\bm V}\ . \label{eq:inputchain}
\end{equation}
Similarly to Antr\`{a}s et al. (cf. Eq. \eqref{eq:up0}), Miller and Temurshoev \cite{Miller} introduced the so-called \emph{downstreamness}, measuring the ``average distance between suppliers of primary inputs and sectors as input purchaser along the input demand supply chain'' as follows:
\begin{equation}
   D_{1i} = 1 \cdot \frac{V_i}{Y_i} + 2\cdot\frac{\sum_{j=1}^N V_j d_{ji}}{Y_i} + 3\cdot \frac{\sum_{j,k=1}^N V_j d_{jk}d_{ki} }{Y_i} + \ldots =   \frac{([\mathds{1}_N - D^T]^{-2}{\bm V})_i}{Y_i} \ .
\end{equation}
As before, using Eq. \eqref{eq:inputchain}, we obtain
\begin{equation}
    {\bm D_1}= [\mathds{1}_N-A_D]^{-1}{\bm 1}_N \ ,\label{eq:D1}
\end{equation}
with \begin{equation}
   A_D= (A Y^{-1})^T = \begin{pmatrix} \frac{a_{11}}{Y_1} & \cdots & \frac{a_{N1}}{Y_1} \\
\vdots& \ddots & \vdots  \\
\frac{a_{1N}}{Y_N} & \cdots & \frac{a_{NN}}{Y_N}  \end{pmatrix}\ . \label{eq:AD}
\end{equation}

The matrix $A_D$ has non-negative elements, and it is row-substochastic, i.e., $\sum_{j}(A_D)_{ij}\leq 1 \ \forall j$.
Finally, as in the upstreamness case, also for the downstreamness, Fally \cite{Fally2012} introduced an analogous iterative definition of the form

\begin{equation}
    D_{2i} = 1 + \sum_{j=1}^N d_{ji}D_{2j} \ ,  \label{eq:D2}
\end{equation}
which can be again mapped with simple manipulations onto Eq. \eqref{eq:D1} using $Y_i d_{ji}=a_{ji}$.

\section{Rank-1 approximation with local and aggregate information}\label{sec:theoframework}
In this section, we will discuss how to derive an approximation for the upstreamness and downstreamness metrics discussed in Sec. \ref
{sec:defUD}.
Let us consider the resolvent $G(A)=(\mathds{1}_N - A)^{-1}$, where the matrix $A$ stands for $A_U$ or $A_D$ as defined in the previous section. Therefore, $A$ has non-negative entries, and is sub-stochastic. Recall that the vectors of upstreamness and downstreamness are defined as ${\bm U}_1 =  G(A_U){\bm 1}_N$ and $ {\bm D}_1= G(A_D){\bm 1}_N$, respectively (cf. Eq. \eqref{eq:fally2}, \eqref{eq:D1}).  We are going now to assume that a detailed and accurate knowledge of all the entries of $A$ is {\em not} available. The only  available aggregate information is given by the  $2N$ constants $\bm r=(r_1,\ldots,r_N)$ and $\bm c=(c_1,\ldots,c_N)$, namely the \emph{sums} of the $N$ rows and columns of $A$. This corresponds to knowing only the total intermediate demand per industry and the total value of all inputs required by each industry respectively. This lack of detailed information is actually quite common in supply chain and intrafirm network analysis \cite{bacilieri}, which in turn leads to the need for inference and reconstruction methods to fill the gaps. 

A simple rank-$1$ approximation $\hat A$ for the matrix $A$ is
\begin{equation}
    \hat A=\frac{1}{N}\bm g\bm q^T=
    \begin{pmatrix}
    \frac{g_1 q_1}{N} & \cdots & \frac{g_1 q_N}{N}\\
    \vdots & \ddots & \vdots\\
    \frac{g_Nq_1}{N} & \cdots & \frac{g_Nq_N}{N}
    \end{pmatrix}\ ,\label{avA}
\end{equation}
where the entries of the column vectors $\bm g = (g_1,\ldots,g_N)$ and $\bm q=(q_1,\ldots,q_N)$ are determined imposing the constraint that $A$ and $\hat A$ share the same row and column sums
\begin{align}
r_i= &\sum_j A_{ij}\equiv \frac{\sum_{k} q_k}{N} g_i=\bar q\ g_i\ , \label{eq:fp1}\\
c_j= &\sum_i A_{ij}\equiv \frac{\sum_{k} g_k}{N} q_i=\bar g\ q_j\ . \label{eq:fp2}
\end{align}
This yields eventually the unique matrix
\begin{equation}
    \hat A =\frac{1}{mN}\bm r\bm c^T\label{hatA}
\end{equation}
with $m=\frac{1}{N} \sum_{ij} A_{ij}=\frac{1}{N}\sum_j c_j=\frac{1}{N}\sum_i r_i$. The rank-$1$ matrix $\hat A$ in \eqref{hatA} is the so-called Maximum Entropy reconstructed matrix (see e.g. \cite{newcorr2,maxent2}) subject to the row and column constraints in \eqref{eq:fp1} and \eqref{eq:fp2} (see also \cite{bianconiannealed,statmech3,statmech1,statmech2,Lowrank} for related works). 

If the only information we have is about row sums, then the corresponding rank-$1$ approximation is even simpler
\begin{equation}
   \hat{A} = \begin{pmatrix}
    \frac{r_1}{N} & \cdots & \frac{r_1}{N}\\
    \vdots & \ddots & \vdots\\
    \frac{r_N}{N} & \cdots & \frac{r_N}{N}
    \end{pmatrix} \label{eq:single}\ .
\end{equation}

Clearly, $\hat A$ has a single non-zero, real and positive eigenvalue $\lambda_1=\frac{1}{mN}\sum_j r_j c_j$ (or $\lambda_1=\frac{1}{N}\sum_j r_j$ in the case of only-row constraints) due to the Perron-Frobenius theorem, and $N-1$ zero eigenvalues, therefore we may expect that this approximation will work better the larger the \emph{spectral gap} (or equivalently the smaller the \emph{spectral radius} in the bulk) \footnote{The spectral gap is defined as $\Gamma=\lambda_1-\Xi$, with $\lambda_1$ real and $<1$ being the Perron-Frobenius eigenvalue. The spectral radius is $ \Xi=\max\{|\lambda_2|,\ldots,|\lambda_{N-1}|\}$.} of the original matrix $A$ is \cite{bartolucciranking, MFPTBCCV}. 
The empirical I-O matrices $A_U,A_D$ typically show a large spectral gap, suggesting that the rank-$1$ approximation described in this section should be very effective.

As the empirical I-O matrices $A_U,A_D$ are rather small ($N=35$), it is more informative to look at their spectral radius. In Sec. \ref{sec:results} we
perform a thorough analysis of the spectra of the I-O matrices at the country level and we study how the accuracy of our rank-$1$ formula is related to the spectral radius. We indeed find that there is a clear negative correlation between the two, i.e. the error made using our approximation increases with $\Xi$. This said, even in the worst cases, the relative errors remain fairly negligible and the formulae work very well across the entire dataset.

Employing this rank-$1$ approximation, we can now evaluate the approximate resolvent 
\begin{equation}
    G(\hat A)=(\mathds{1} - \hat A)^{-1}=
    \mathds{1}+\frac{\hat A}{1-\frac{1}{m N}\sum_j r_j c_j}\ ,\label{approxResolvent}
\end{equation}
using the Sherman-Morrison formula \cite{sherman1950} for the inverse of a rank-$1$ matrix, from which it follows that the upstreamness and downstreamness of the $i$-th industry are respectively approximated by
\begin{align}
    U_{1i} &\approx 1+\frac{r_i}{1-\frac{1}{m N}\sum_j r_j c_j}\label{eq:approxU}\\
    D_{1i} &\approx 1+\frac{\tilde r_i}{1-\frac{1}{\tilde m N}\sum_j \tilde r_j \tilde c_j}\label{eq:approxD}\ ,
\end{align}
where $r_i,c_i$ and $\tilde r_i,\tilde c_i$ represent respectively the sum of rows and columns of $A_U$ and $A_D$. If only the constraint on row is imposed, the formulae above reduce to

\begin{align}
    U_{1i} &\approx 1+\frac{r_i}{1-\frac{1}{N}\sum_j r_j }\label{eq:approxUsingle}\\
    D_{1i} &\approx 1+\frac{\tilde r_i}{1-\frac{1}{ N}\sum_j \tilde r_j }\label{eq:approxDsingle}\ .
\end{align}

The approximate formulae above show that, within our rank-$1$ approximation, the upstreamness (downstreamness) of sector $i$ is fully determined by the interplay of (i) \emph{local} and \emph{aggregate} information, namely of the total intermediate demand per sector (and/or the total value of all inputs required by a each sector), and (ii) a suitable average of the total intermediate demand (and/or the total value of all inputs) across \emph{all} sectors in the economy. 

In spite of the seemingly drastic approximation, which neglects a significant amount of finer intersectorial details, we will show that the aggregate information featuring in our rank-1 formulae is sufficient to determine with high accuracy the relative positioning of countries and sectors within the global value chains. 

In the next sections, we will then calculate upstreamness and downstreamness measures on I-O tables from the NIOT Dataset (see Sec. \ref{Dataset}), comparing the results obtained via our approximation with the full calculation using the original formulae, namely Eq. \eqref{eq:fally2} and \eqref{eq:D1}.

\section{Dataset}\label{Dataset}
\begin{table}[htb!]
\centering
\footnotesize
\begin{tabular}{c|c |c}
\hline
Australia (AUS)     & France (FRA)          & Netherland (NLD)\\ 
Austria (AUT)       & Great Britain (GBR)   & Poland (POL)\\
Belgium  (BEL)      & Greece (GRC)          & Portugal (PRT)\\
Bulgaria  (BGR)     & Hungary (HUN)         & Romania (ROU) \\
Brazil (BRA)        & Indonesia (IDN)       & Russia (RUS)\\
Canada  (CAN)       & India (IND)           & Slovakia (SVK)\\
China (CHN)         & Ireland (IRE)         & Slovenia (SVN)\\
Cyprus (CYP)        & Italy (ITA)           & Sweden (SWE)\\
Czech Republic (CZE)& Japan  (JPN)          & Turkey (TUR)\\
Germany (DEU)       & Korea (KOR)           & Taiwan (TWN) \\
Denmark (DNK)       & Lituania (LTU)        & United States (USA) \\
Spain (ESP)         & Latvia (LVA)          & Luxembourg (LUX)$^\star$\\
Estonia (EST)       & Mexico (MEX)          &\\
Finland (FIN)       & Malta (MLT)           &\\
\hline
\end{tabular}
\caption{Countries and their codes in the NIOT database by WIOD \cite{wiotdataset}. Luxembourg is not included in our analysis as data present inconsistencies across the years.}
\label{tab:countries}
\end{table}
The empirical I-O matrices used for the experiments have been constructed using the 2013 release of the National Input-Output tables by the World Input-Output Database (WIOD) \cite{wiotdataset}. 
The NIOT dataset comprises 39 countries --representing a large fraction of the major world economies -- over the years 1995 - 2011. The list of countries and their codes considered in our empirical analysis is presented in Tab. \ref{tab:countries}. 
The structure of the input-output table of each country is schematically shown in Fig. \ref{fig:ioscheme2}. The intermediate demand for each country is reported for $N=35$ economic sectors in terms of the flow (in US million dollars) between sectors. The full list of economic sectors and their codes included in our analysis is summarized in Tab. \ref{tab:sectors}.
The final demand is characterized in terms of (i) final consumption expenditure by households, (ii) final consumption expenditure by non-profit organizations serving households (NPISH), (iii) final consumption expenditure by government,  (iv) gross fixed capital formation, (v) changes in inventories and valuables and (vi) exports. 
In the dataset sometimes the change in Inventories and Valuables can be negative, and were assumed to contribute to imports.
The entries $a_{ij}$ of each row of the full I-O table are then normalized by the vector outputs $Y_j$. The normalized intermediate demand sub-matrix is sub-stochastic and represents the matrix $A_U$. The $r_i$ used in the model are simply the sums over the rows of the matrix $A_U$ (or equivalently if normalized by columns the matrix $A_D$, respectively in Eq.~\eqref{eq:AU} and \eqref{eq:AD}) .

\begin{table}[H]
\footnotesize
    \centering
    \begin{tabular}{c|c}
    \hline
Agriculture, Hunting, Forestry and Fishing & AGR \\
Mining and Quarrying & MIN\\
Food, Beverages and Tobacco & FOD \\
Textiles and Textile Products & TXT \\
Leather, Leather and Footwear & LEA \\
Wood and Products of Wood and Cork & WOO \\
Pulp, Paper, Paper , Printing and Publishing & PRT \\
Coke, Refined Petroleum and Nuclear Fuel & COK \\
Chemicals and Chemical Products & CHM \\
Rubber and Plastics & RUB \\
Other Non-Metallic Mineral & NMM \\
Basic Metals and Fabricated Metal & MET \\
Machinery, Nec & MAC \\
Electrical and Optical Equipment & ELO \\
Transport Equipment & TRA \\
Manufacturing, Nec; Recycling & MAN \\
Electricity, Gas and Water Supply & ELE \\
Construction & CON \\
Sale, Maintenance and Repair of Motor Vehicles and Motorcycles; Retail Sale of Fuel & MOT \\
Wholesale Trade and Commission Trade, Except of Motor Vehicles and Motorcycles & WHO \\
Retail Trade, Except of Motor Vehicles and Motorcycles; Repair of Household Goods & RET \\
Hotels and Restaurants  & HOT \\
Inland Transport &ITR \\
Water Transport & WAT \\
Air Transport & AIR \\
Other Supporting and Auxiliary Transport Activities; Activities of Travel Agencies & OTR \\
Post and Telecommunications & POS \\
Financial Intermediation & FIN \\
Real Estate Activities & EST \\
Renting of MEq and Other Business Activities & REN \\
 Public Admin and Defence; Compulsory Social Security & PUB \\
Education & EDU \\
Health and Social Work & HEA \\
Other Community, Social and Personal Services & SOC \\
Private Households with Employed Persons & HOU\\
\hline
    \end{tabular}
    \caption{Sectors of the NIOT dataset by WIOD (2013 release) and their sector codes \cite{wiotdataset}.}
    \label{tab:sectors}
\end{table}  
\section{Results} \label{sec:results}
In this section, we compare our approximate formulae for downstreamness and upstreamness with single (Eq. \eqref{eq:approxUsingle} and \eqref{eq:approxDsingle} respectively) and double contraints (Eq. \eqref{eq:approxU} and \eqref{eq:approxD} respectively) with the measures obtained via direct inversion of the empirical I-0 matrix (Eq. \eqref{eq:fally2} and \eqref{eq:D1} respectively).

\begin{figure}[H]
    \centering
    \includegraphics[width=0.80\textwidth]{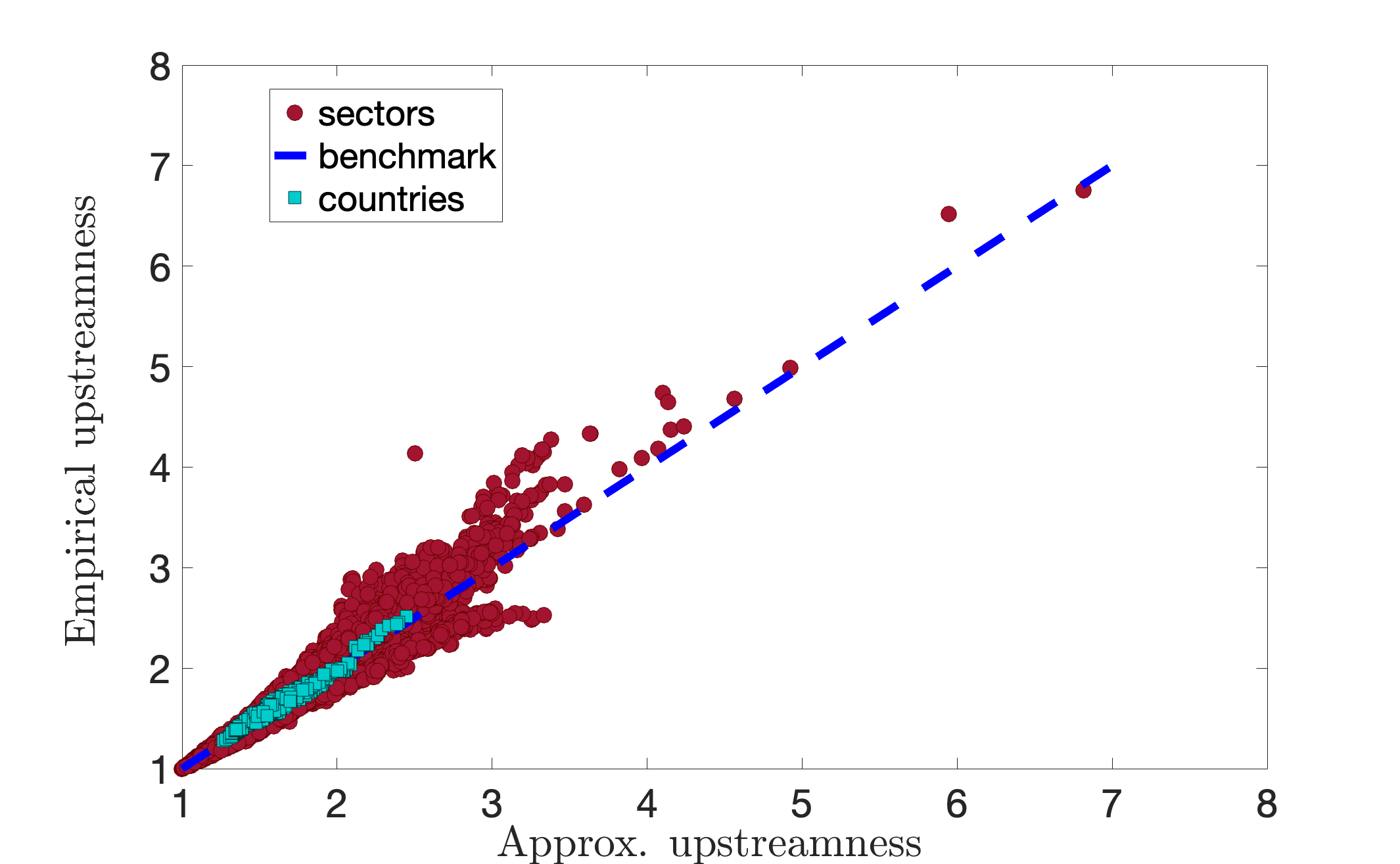}
    \includegraphics[width=0.80\textwidth]{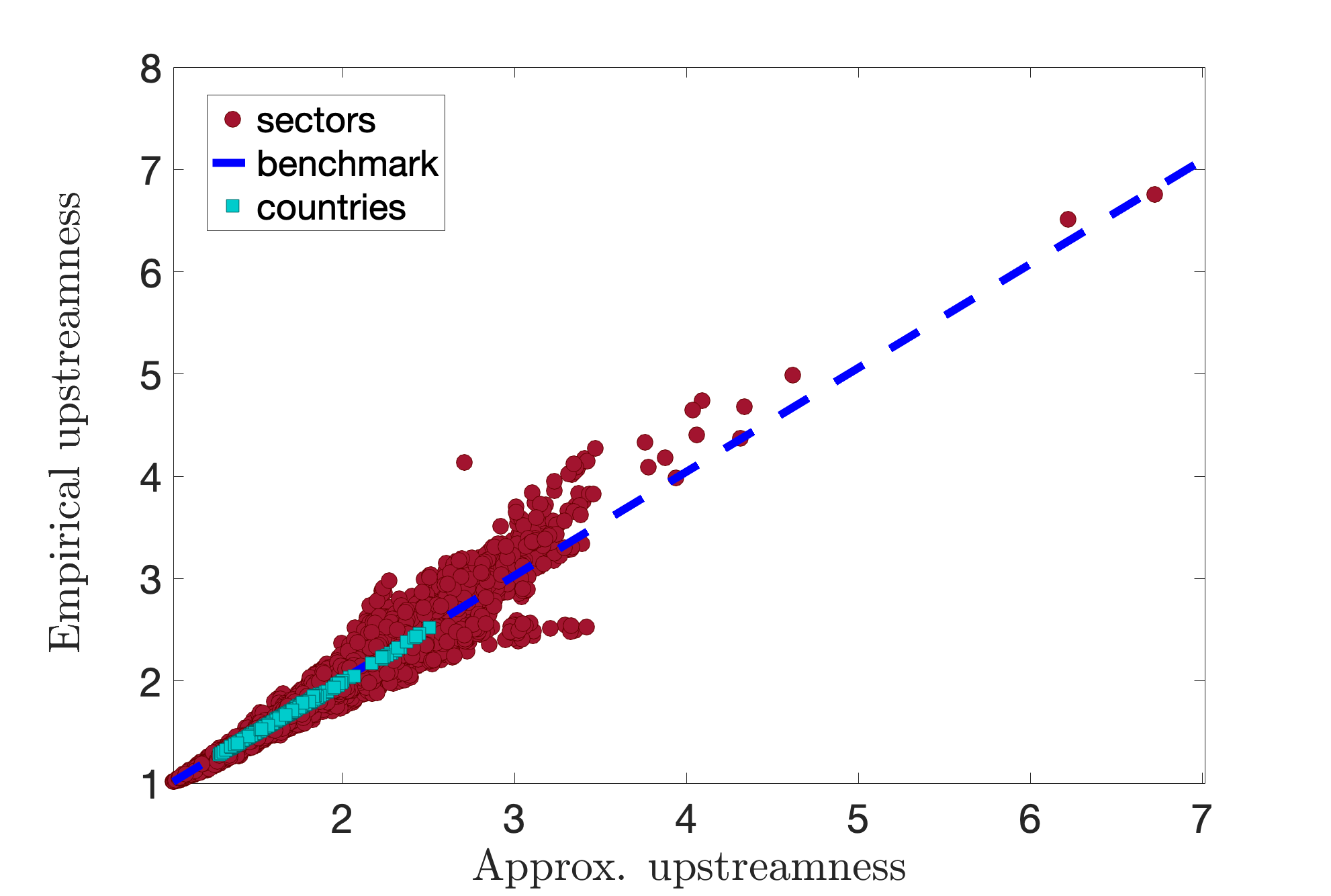}

    \caption{Empirical upstreamness versus approximated upstreamness. Cyan squares represent the upstreamness per country ($39$  countries) per year ($11$ year) averaged over $35$ industrial sectors from the WIOD dataset (Release 2013). Red full circles represents the upstreamness for all industry sectors in all countries/all years. Top panel: Empirical upstreamness compared with single-constraint approximation in Eq. \eqref{eq:approxUsingle}.
    Bottom panel: Empirical upstreamness compared with double-constraints approximation in Eq. \eqref{eq:approxU}.}
    \label{fig:upstream}
\end{figure}

\begin{figure}[H]
    \centering
    \includegraphics[width=0.80\textwidth]{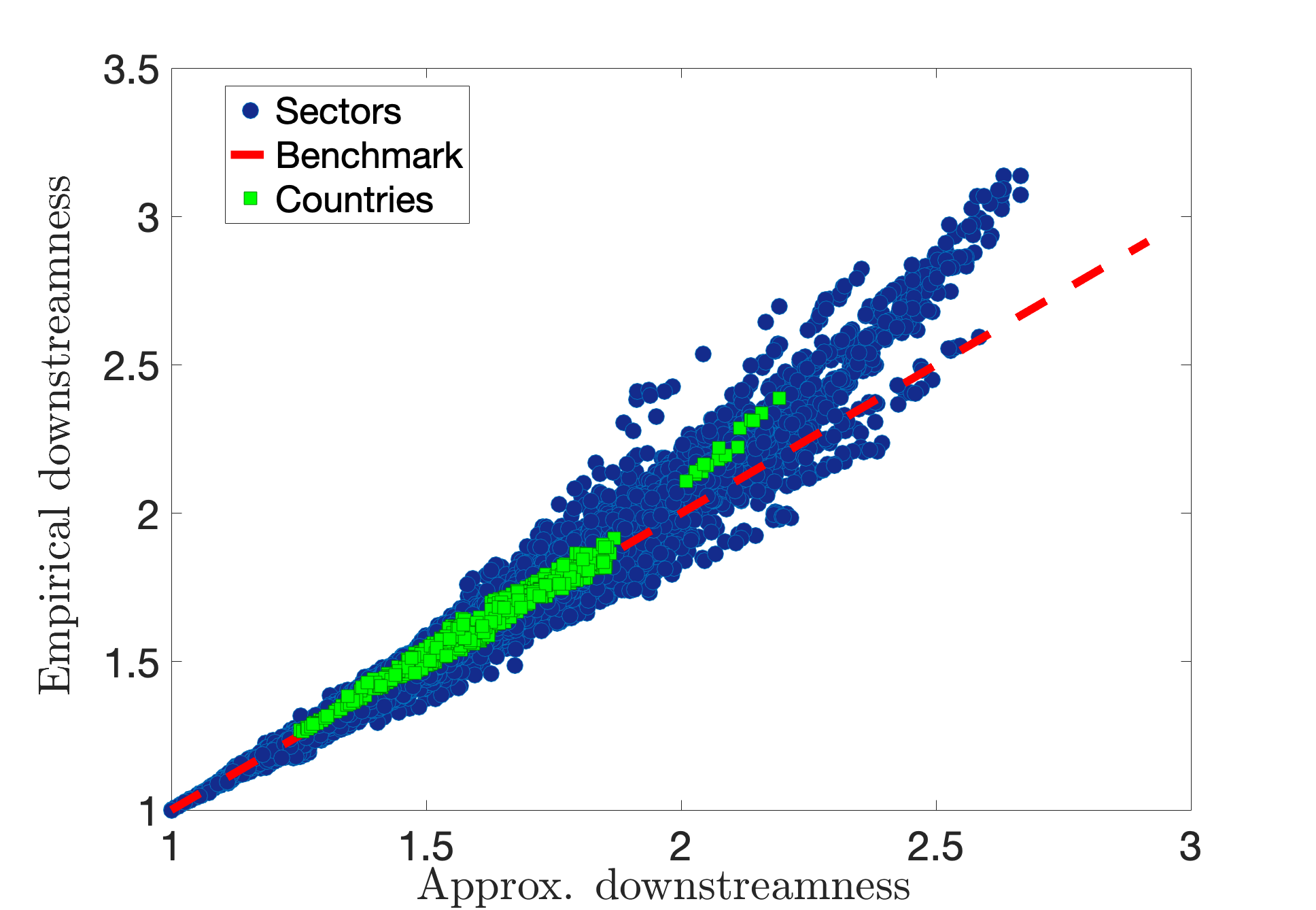}
    \includegraphics[width=0.80\textwidth]{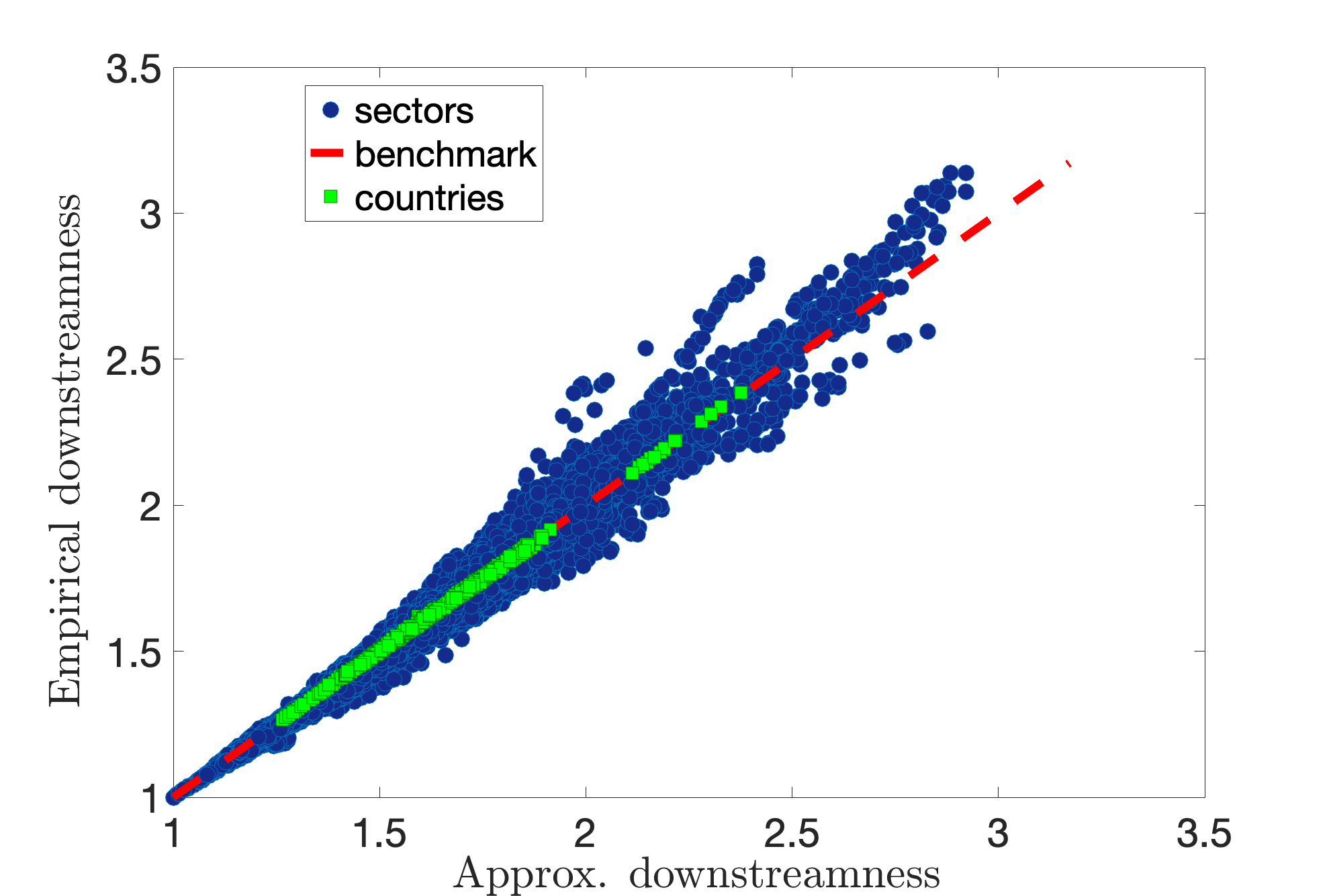}

    \caption{Empirical downstreamness versus approximated downstreamness. Light green squares represent the downstreamness per country ($39$  countries) per year ($11$ year) averaged over $35$ industrial sectors from the WIOD dataset (Release 2013).  Blue full circles represent the downstreamness of all industry sectors in all countries/all years. Top panel: Empirical upstreamness compared with single-constraint approximation in Eq. \eqref{eq:approxDsingle}.
    Bottom panel: Empirical upstreamness compared with double-constraints approximation in Eq. \eqref{eq:approxD}. }
    \label{fig:downstream}
\end{figure} 
In Fig. \ref{fig:upstream} we plot the empirical average over all sectors (cyan squares) of the \textit{upstreamness} for $39$ countries (listed in Table \ref{tab:countries}) for all years (1995-2011) versus the approximate value with single (top panel) and double constraints (bottom panel), respectively obtained in Eq. \eqref{eq:approxUsingle} and Eq. \eqref{eq:approxU}. 
We see that the empirical data ($663$ data points - $39$ countries $\times$ $17$ years) nicely collapse on top of the theoretical benchmark (blue dashed line). In the single constraint case, this implies that the average upstreamness coefficient for a country is determined with high accuracy by the knowledge of a single quantity $\bar{z} = 1 - \frac{1}{N}\sum_j{r_j}$, corresponding to one minus the average total intermediate demand.
We also show the upstreamness values for each sector in each country across the entire period (red full circles) constituting in total $\sim23k$ data points - $35$ sectors $\times$ $39$ countries $\times$ $17$ years. At the sector level, we observe a similar good agreement of the empirical exact upstreamness with the approximate values.
 There are occasional deviation, whose origin can be traced back to a higher degree of heterogeneity in the $A$ matrix with respect to the ``flat" rank-1 model introduced in Eq. \eqref{eq:single}. In the following we will also analyze more closely the relation between the error -- discrepancy between the actual values of upstreamness (and downstreamness) calculated via direct inversion and those obtained via our approximate formula -- and the spectral properties of the empirical I-O matrix $A$. 

\begin{figure}[H]
    \centering
    \includegraphics[width=0.8\textwidth]{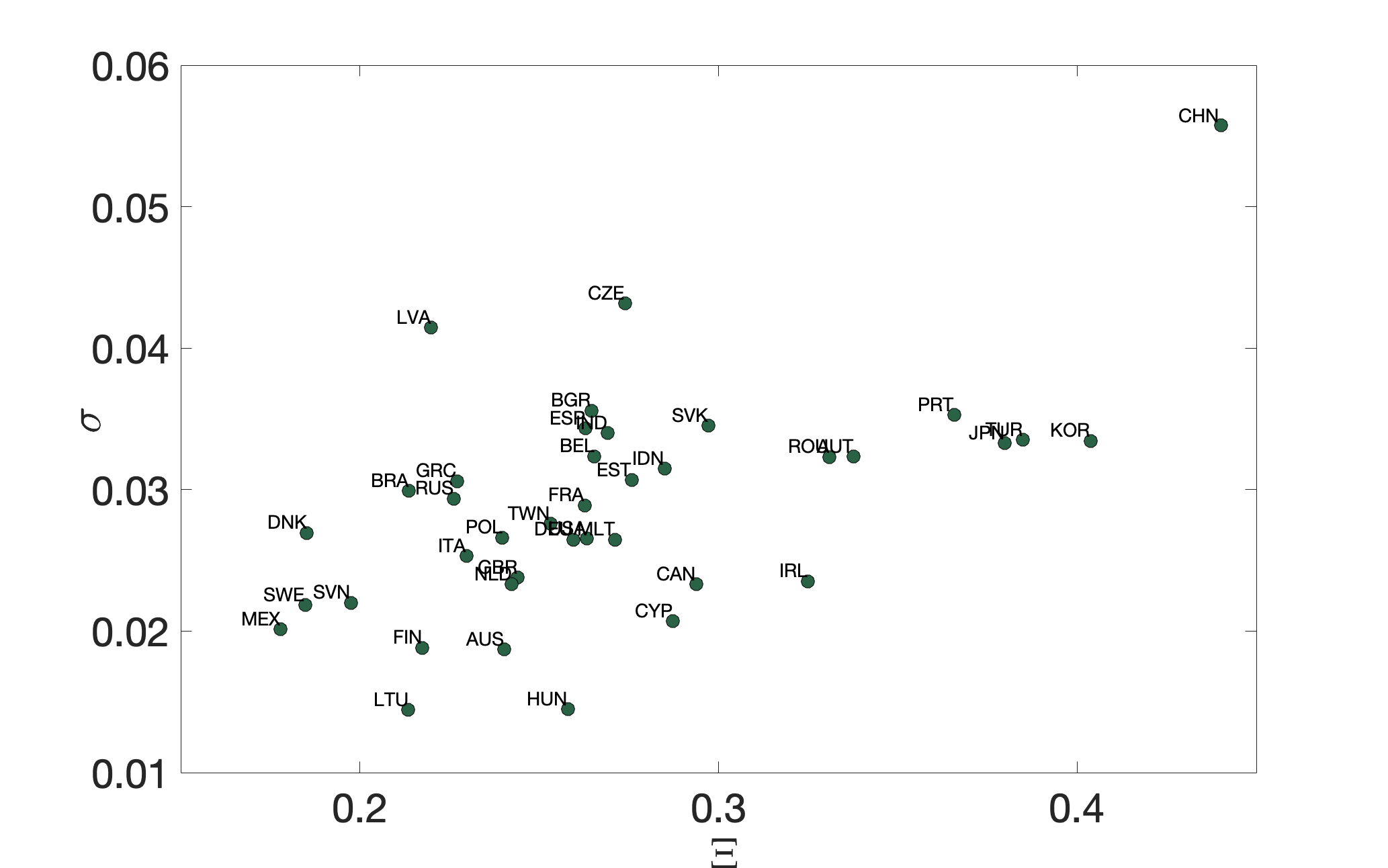}
    \includegraphics[width=0.8\textwidth]{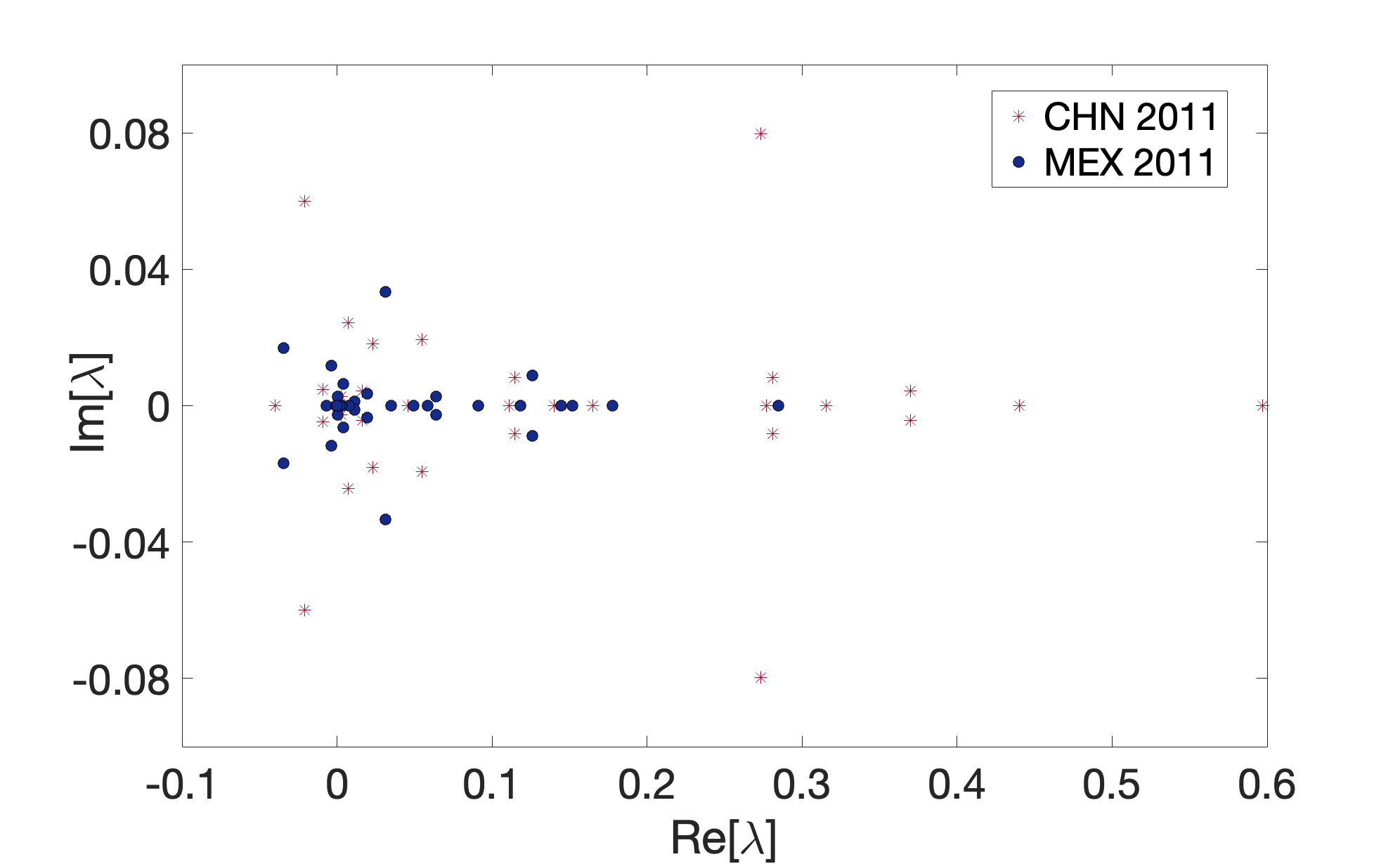}
    \caption{Top panel: Error $\sigma$  approximated vs. exact upstreamness averaged over the sectors as a function of the spectral radius $\Xi$ of the matrix $A_U$ for all 39 countries in 2011. Bottom panel: Eigenvalue spectrum of the $A_U$ matrix of China (CHN) and Mexico (MEX) in 2011.}
    \label{fig:spectrum2011}
\end{figure}

In Fig. \ref{fig:downstream}, we repeat a similar analysis for the \textit{downstreamness}, comparing the values obtained via direct inversion (Eq. \eqref{eq:D1}) with the approximate values of downstreamness imposing the single or double constraint on the knowledge of row sums, or row and column sums, respectively. Also for this measure, we observe a good agreement between exact and approximate values, both at the sectors (red full circles) and at the aggregate country level (cyan squares).
\begin{figure}[htb!]
    \centering
    \includegraphics[width=0.8\textwidth]{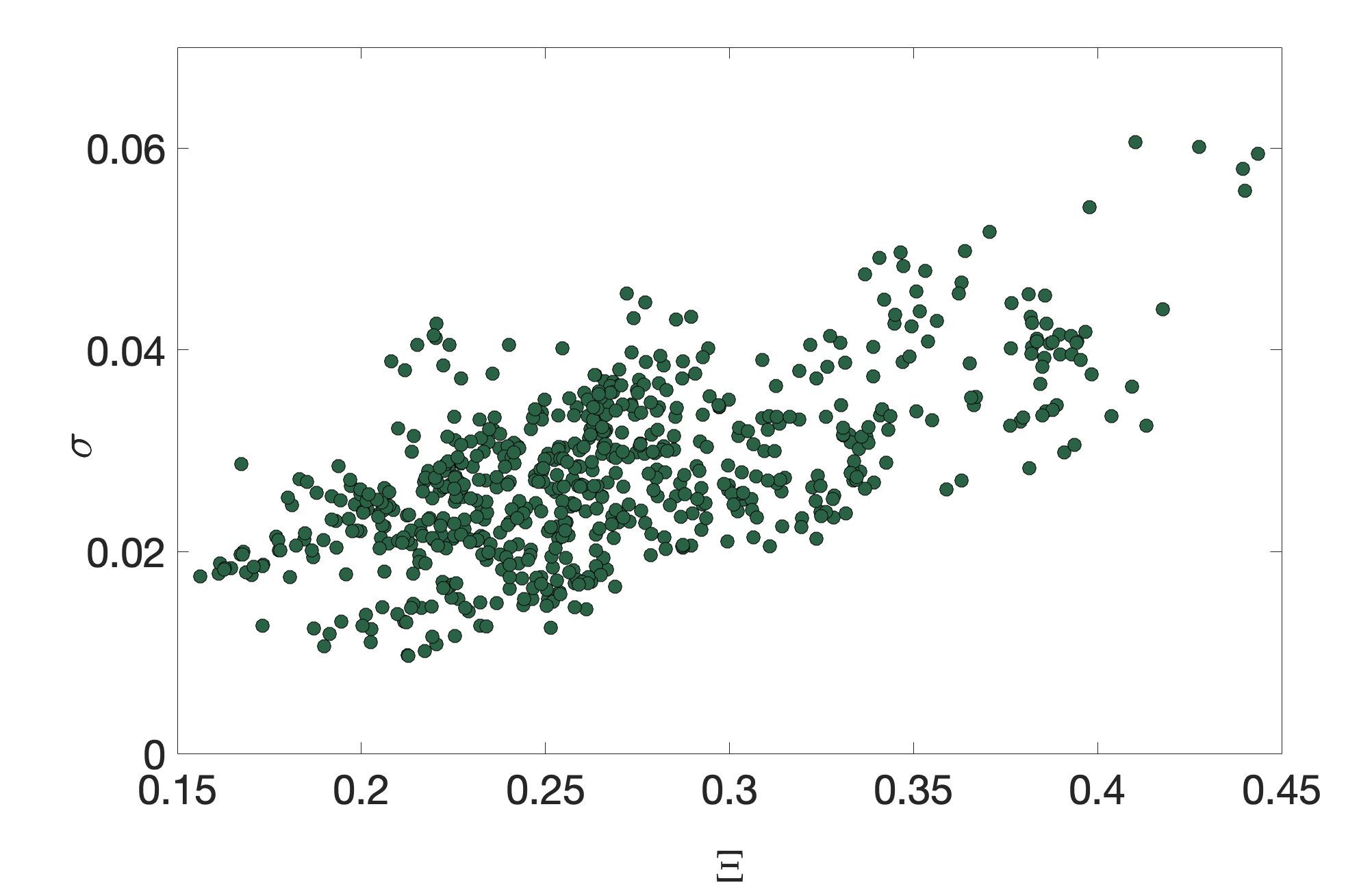}
    \caption{Error $\sigma$ on approximated vs. exact upstreamness calculated for $39$ countries, for the years $1995-2011$ year averaged over the sectors as a function of the spectral radius $\Xi$.}
    \label{fig:radiusall}
\end{figure}
In the following, we analyze more closely the error made in the estimation of the upstreamness/downstreamness coefficients via our approximate formulae and link it to spectral properties of the underlying I-O matrix $A$.
In particular, we define the following metric for assessing the error \cite{bartolucciranking}
\begin{equation}
\sigma=\left\langle\left|\frac{\mathcal{R}_i^{(\rm{emp})}}{\mathcal{R}_i^{(\rm{approx})}}-1\right|\right\rangle \ , 
\label{eq:sigma}
\end{equation}
where $\mathcal{R}_i$ represents either the upstreamness or the downstreamness values computed via direct inversion ($\mathcal{R}_i^{(\rm{emp})}$) and via our approximate formula ($\mathcal{R}_i^{(\rm{approx})}$) respectively. The average $\langle\cdots\rangle$ is calculated over all sectors of a given country.
Concerning the spectral properties, as shown in \cite{bartolucciranking,MFPTBCCV} the accuracy of the approximation is related to the spectral gap of the matrix $A$. The matrix $A$ has non-negative entries, therefore it has one real eigenvalue of largest magnitude $\lambda_1$ (the Perron-Frobenius eigenvalue), and its spectral gap is defined as $\Gamma=\lambda_1-\max\{|\lambda_2|,\ldots,|\lambda_{N-1}|\}$. As the empirical I-O matrices are rather small ($N=35$) it is more informative to look at the spectral radius.
We then introduce the spectral radius excluding the Perron-Frobenius $\lambda_1$ as
\begin{equation}
    \Xi=\max\{|\lambda_2|,\ldots,|\lambda_{N-1}|\}\  . \label{eq:radius}
\end{equation}
This definition is consistent with the approach used in the case of Gaussian matrices perturbed with a rank-1 matrix that may force an outlier to split off from the circular bulk \cite{MFPTBCCV,DeGiuli}. 
In Fig. \ref{fig:radiusall}, we display the error $\sigma$ made on the approximation for all countries in all years as a function of the spectral radius $\Xi$ of the $A_U$ matrix characterizing each country in each year. As expected, the error grows with the spectral radius, as the rank-1 approximation becomes less accurate in reproducing the underlying intersectorial interactions. In Fig. \ref{fig:spectrum2011}, we show the same relationship labelling the countries for a single year (2011). In the bottom panel, we show the eigenvalue spectrum of two selected countries -- namely China and Mexico -- displaying respectively among the maximal and minimal errors in the estimation, to highlight spectral differences in the displacement of eigenvalues in the bulk.
In this analysis, we find a clear negative correlation between the accuracy of the estimation and the spectral radius, i.e. the error made using our approximation increases (equivalently the accuracy of the approximation decreases) with $\Xi$. In general though, even in the worst cases, the relative errors remain fairly small ($\sim 5-6\%$) and the approximation works very well across the entire sample.

\section{Upstreamness under aggregation}\label{sec:aggregation}

In this section, we briefly consider how our approximation performs after the Input-Output data matrix has been subject to aggregation (consolidation) of different industrial sectors. The effects of aggregation -- i.e. the procedure by which the data are looked at and lumped together at different ``granularity'' level -- have been considered in many works (see \cite{AggregationReview} for a comprehensive review). Here we consider the axiomatic formulation of aggregation provided in \cite{axiomatic}, which is summarized below. Furthermore, our treatment will be confined to the upstreamness, and the row-only rank-$1$ approximation, as generalizations to the other cases are straightforward. 

Consider the definition of upstreamness given in Eq. \eqref{eq:up1}
\begin{equation}
  {\bm U_1}= [\mathds{1}_N-A_U]^{-1}{\bm 1}_N\ .
        \label{eq:fally2aggr}
\end{equation}
To make contact with Ref. \cite{axiomatic}, we rewrite \eqref{eq:fally2aggr} as
\begin{equation}
  [{\bm U_1}^T]_N= {\bm 1}_N^T[\mathds{1}_N-A_U^T]^{-1}\ ,
        \label{eq:fally2aggr2}
\end{equation}
in terms of row vectors $ {\bm U_1}^T$ and ${\bm 1}_N^T$, and a \emph{column}-substochastic $N\times N$ matrix $A_U^T$. The notation $[\ldots]_N$ indicates that the vector has length $N$.

Let us assume that we wish to aggregate the $N$ ``micro'' industrial sectors or commodities into a set of $M<N$ ``macro'' sectors or commodities. Formally, we can define two matrices, $S$ and $T$, of size $M\times N$ and $N\times M$ respectively. The $\{0,1\}$ matrix $S$ indicates which micro-sectors should be combined together: $S_{ij}=1$ if micro-sector $j$ is to be included in macro-sector $i$. Thus, $S$ is a column stochastic matrix with exactly one $1$ in every column, and at least one $1$ in every row. The matrix $T$ indicates the proportional weights of each micro-sector within its macro-aggregate. The element $T_{ji}\in (0,1)$ represents the weight $w_{ji}$ that micro-sector $j$ carries within macro-sector $i$, and therefore is such that $\sum_j T_{ji}=1$. It follows that $T$ is also column stochastic. 

Forming the aggregate $M\times M$ matrix $A_U^\prime=SA_U^T T$ is the most common way used in the literature to create a smaller sub-stochastic matrix from the original matrix $A_U$, which retains (at a coarser level of detail) some of the information about industrial sectors and commodities provided by $A_U$. Although other choices of aggregation are possible, it was proven in \cite{axiomatic} that the aggregator $A_U^\prime$ is the only one that satisfies three natural axioms of \emph{linearity}, \emph{value added neutrality}, and \emph{partitioning}, therefore in the following we will confine ourselves to this case (the so called \emph{standard aggregator}). It follows from the definition of $S$ and $T$ that $ST=\mathds{1}_M$ and $TS$ is a column stochastic, idempotent matrix of rank $M$ (see \cite{axiomatic} for a proof). 

Although in principle any non-negative column-stochastic matrix could play the role of $T$, in practice it makes most sense to define it as
\begin{equation}
    T=\mathrm{diag}(\bm w)S^T [\mathrm{diag}(S\bm w)]^{-1}\ ,
\end{equation}
where $\bm w$ is a vector of $N$ non-negative numbers, and $\mathrm{diag}(\bm w)$ is the diagonal matrix having the vector entries on the diagonal (in their natural order). According to Charnes and Cooper, ``The main justification for this mode of consolidation is that it conforms to the way data would be synthesized ab initio if $SAT$ rather than $A$ were the objective" \cite{charnes}. To better understand how standard aggregation works, consider as an example a $6\times 6$ matrix $A_U^T$ (whose elements we denote $\alpha_{ij}$ for simplicity, so $\alpha_{ij} = a_{ji}/Y_j$). Let 
\begin{equation}
   S = \begin{pmatrix}
    0 & 0 & 1 & 1 & 0 & 0\\
    1 & 1 & 0 & 0 & 0 & 0\\
    0 & 0 & 0 & 0 & 1 & 1\\
    \end{pmatrix} \label{eq:Saggregator}\ ,
\end{equation}
and $\bm w = (w_1,w_2,w_3,w_4,w_5,w_6)$. Then
\begin{equation}
   T = \mathrm{diag}(\bm w)S^T [\mathrm{diag}(S\bm w)]^{-1}=
\begin{pmatrix}
 0 & \frac{w_1}{w_1+w_2} & 0 \\
 0 & \frac{w_2}{w_1+w_2} & 0 \\
 \frac{w_3}{w_3+w_4} & 0 & 0 \\
 \frac{w_4}{w_3+w_4} & 0 & 0 \\
 0 & 0 & \frac{w_5}{w_5+w_6} \\
 0 & 0 & \frac{w_6}{w_5+w_6} \\
\end{pmatrix}
 \label{eq:Taggregator}\ ,
\end{equation}
and the aggregator becomes
\begin{equation}
    A_U^\prime=S A_U^T T =
\begin{pmatrix}
 \frac{w_3 (\alpha_{33}+\alpha_{43})+w_4 (\alpha_{34}+\alpha_{44})}{w_3+w_4} & \frac{w_1 (\alpha_{31}+\alpha_{41})+w_2 (\alpha_{32}+\alpha_{42})}{w_1+w_2} & \frac{w_5 (\alpha_{35}+\alpha_{45})+w_6 (\alpha_{36}+\alpha_{46})}{w_5+w_6} \\
 \frac{w_3 (\alpha_{13}+\alpha_{23})+w_4 (\alpha_{14}+\alpha_{24})}{w_3+w_4} & \frac{w_1 (\alpha_{11}+\alpha_{21})+w_2 (\alpha_{12}+\alpha_{22})}{w_1+w_2} & \frac{w_5 (\alpha_{15}+\alpha_{25})+w_6 (\alpha_{16}+\alpha_{26})}{w_5+w_6} \\
 \frac{w_3 (\alpha_{53}+\alpha_{63})+w_4 (\alpha_{54}+\alpha_{64})}{w_3+w_4} & \frac{w_1 (\alpha_{51}+\alpha_{61})+w_2 (\alpha_{52}+\alpha_{62})}{w_1+w_2} & \frac{w_5 (\alpha_{55}+\alpha_{65})+w_6 (\alpha_{56}+\alpha_{66})}{w_5+w_6} \\
\end{pmatrix}\ .
\end{equation}

Now, let us assume that the vector of $N$ upstreamness values in \eqref{eq:fally2aggr2} can be faithfully approximated by our formula \eqref{eq:approxUsingle}, which can be written as
\begin{equation}
  [\hat{\bm U_1}^T]_N= {\bm 1}_N^T+\frac{1}{1-\bar{r}_N}\bm r^T\ ,
        \label{eq:fally2aggr2approx}
\end{equation}
where $\bm r$ is the (column) vector of row sums of the matrix $A_U$ (or the column sums of $A_U^T$, $r_j = \sum_{i=1}^N \alpha_{ij}$), and $\bar r_N$ is their average. Let us further assume that the original data matrix $A_U$ is not known in its entirety (only its row sums are known), but the sectors/commodities in $A_U$ have been aggregated using a \emph{known} pair of matrices $S,T$ -- in other words, we are aware of what sectors/commodities have been lumped together (and with which relative weights) and what their aggregate outputs are, but we do not have more detailed information. We ask whether the knowledge of $\bm r, S$ and $T$ is sufficient to determine $[\hat{\bm U_1}^T]_M$, namely a faithful approximation for the $M$ upstreamness values of the aggregate model. The answer is affirmative. 

First, define
\begin{equation}
  [{\bm U_1}^T]_M= {\bm 1}_M^T[\mathds{1}_M-A_U^\prime]^{-1}={\bm 1}_M^T[\mathds{1}_M-SA_U^T T]^{-1}\ ,
        \label{aggregatedU}
\end{equation}
the vector of $M$ upstreamness values, obtained using the aggregate matrix $A_U^\prime$ as a source. The Leontief matrix on the r.h.s. of \eqref{aggregatedU} is equal to the aggregate of the Leontief matrix of the so called \emph{companion matrix} $\bar A_U= A_U^T TS$ \cite{axiomatic}, namely\footnote{The proof follows by expanding $[\mathds{1}_M-SA_U^T T]^{-1}=\mathds{1}_M +SA_U^T T+(SA_U^T T)^2+\ldots$, and using $(SA_U^T T)^n=S(A_U^T TS)^nT$ and $TST=T$.}
\begin{equation}
  [\mathds{1}_M-SA_U^T T]^{-1} =  S[\mathds{1}_N-\bar A_U]^{-1}T\ .\label{theoremFeicompanion}
\end{equation}

Imagine now that the true matrix $A_U^T$ appearing on the l.h.s. of \eqref{theoremFeicompanion} is replaced by its best rank-$1$ approximation, given by $\hat A^T$ (see Eq. \eqref{eq:single}). It is easy to deduce\footnote{This follows from the fact that the rank of the product of two matrices ($\hat A$ and $TS$) is smaller or equal than the smallest rank of the two factors, and that $TS$ is rank-$M$ (and of course none of the matrices involved is a null matrix).} that in this case the companion matrix will also be rank-$1$. Applying Sherman-Morrison on the r.h.s. of \eqref{theoremFeicompanion}, we get
\begin{equation}
    S[\mathds{1}_N-\hat A TS]^{-1}T=
    \mathds{1}_M+\frac{1}{1-\phi(\bm r,S,T)}S(\hat A TS)T=\mathds{1}_M+\frac{1}{1-\phi(\bm r,S,T)}S \hat A T\ ,\label{eq:finalaggr}
\end{equation}
where we used $S\mathds{1}_N T=ST=\mathds{1}_M$, and
\begin{equation}
    \phi(\bm r,S,T)=\frac{1}{N}\sum_{i,k=1}^N r_i (TS)_{ik}\ .
\end{equation}
Eq. \eqref{eq:finalaggr} shows how to construct a faithful rank-$1$ approximation for the upstreamness of the aggregate model starting from the knowledge of row sums of the original model, as well as of the matrices $T$ and $S$ implementing the aggregation.

\section{Summary and Outlook}\label{sec:conclusions}

In this paper, we have shown that the upstreamness and downstreamness measures introduced in the context of Input-Output analysis at both the inter-sectorial and country level can be faithfully recovered from the knowledge of \emph{aggregate} and \emph{local} information about the I-O table. In other words, the precise determination of the elements of the input-output matrix does not matter much, as long as their distribution does not deviate significantly from the ``homogeneous" (flat) model (described in Eq. \eqref{eq:single}), and the total intermediate demand per sector is ordinarily sufficient to provide an accurate estimate of the sector's multipliers. 

Our rank-1 approximation has been successfully tested on National Input-Output tables obtained from WIOD, where an excellent correlation is obtained between the empirical multipliers and the theoretical formulae (see Fig. \ref{fig:upstream} and Fig. \ref{fig:downstream}). Small deviations from this remarkably robust regularity are readily attributed to stronger heterogeneity in the empirical sectorial data, which would require refinements to the (single or doubly constrained) rank-$1$ approximation presented here.

Indeed, sparser or more heterogeneous I-O matrices tend to have a larger spectral radius (or equivalently a smaller spectral gap), as demonstrated in Fig. \ref{fig:radiusall} and Fig. \ref{fig:spectrum2011}. The quality of our rank-$1$ approximation is very high across the sectors and countries considered, but may be inferior for emprical matrices larger spectral radii -- as more eigenvalues besides the largest (Perron-Frobenius) start to play an important role. 

In Section \ref{sec:aggregation}, we have also shown how our rank-$1$ approximation is well-behaved with respect to aggregation of sectorial data: knowing what sectors/commodities are lumped together, and what their aggregate outputs are, is sufficient to determine a faithful approximation for the upstreamness values of the aggregate model, as the rank-$1$ nature of the approximation is preserved upon aggregation.

In a recent paper \cite{puzzling}, we further employ the rank-$1$ approximation as a proxy to investigate the ``puzzling'' correlations observed between upstreamness and downstreamness at aggregate level \cite{Antras2018}. More generally, our approach based on a rank-$1$ approximation demonstrates that local and aggregate information about I-O tables is ordinarily sufficient to determine the upstreamness and downstreamness at sectorial and country level with high accuracy, while at the same time providing analytically tractable formulae (Eq.~\eqref{eq:D1}, \eqref{eq:up1}) that avoid matrix inversions altogether. As an outlook for future research, it will be interesting to test the accuracy of our formulae on firm-level data, where data availability and sparsity are greater concerns. In spite of the sparser nature of the data, we would expect our approximation to work well, as recently shown on experiments conducted on synthetic data \cite{bartolucciranking}.

\subsection*{Acknowledgements} We gratefully acknowledge insightful conversations with J. D. Farmer, F. Lafond, L. P. Garcia-Pinto and J. McNerney.

\end{document}